\documentclass[preprint,prd,nofootinbib,onecolumn,showpacs]{revtex4}
\usepackage{dcolumn}
\usepackage{lipsum}
\usepackage{graphicx,epsfig}
\usepackage{psfrag}
\usepackage{bm}
\usepackage{epstopdf}
\usepackage{latexsym}
\usepackage{multirow}
\usepackage{amsmath,amssymb}
\usepackage{enumerate}
\usepackage{hyperref}
\usepackage[FIGTOPCAP]{subfigure}

\def\be {\begin{equation}}
\def\ee {\end{equation}}
\def\ba {\begin{eqnarray}}
\def\ea {\end{eqnarray}}
\def\nn {\nonumber}
\def\bc {\begin{center}}
\def\ec {\end{center}}
\newcommand{\bdm}{\begin{displaymath}}
\newcommand{\edm}{\end{displaymath}}

\usepackage{multirow}
\usepackage{amsmath,amssymb}
\usepackage{enumerate}
\usepackage{hyperref}

\def\be {\begin{equation}}
\def\ee {\end{equation}}
\def\ba {\begin{eqnarray}}
\def\ea {\end{eqnarray}}
\def\bc {\begin{center}}
\def\ec {\end{center}}
\def\bi {\begin{itemize}}
\def\ei {\end{itemize}}

\DeclareMathOperator{\sech}{sech}
\DeclareMathOperator{\F}{F} % elliptic F
\DeclareMathOperator{\E}{E}  % elliptic E
\def\a  {\alpha}
\def\b  {\beta}

\def\d  {\delta}

\def\da {\dagger}

\def\f {\frac}
\def\g  {\gamma}
\def\G  {\Gamma}

\def\hb {\hbar}

\def\l  {\lambda}
\def\la {\label}
\def\le {\left}

\def\m  {\mu}

\def\na {\nabla}
\def\nn {\nonumber}

\def\o  {\omega}
\def\O  {\Omega}

\def\pa {\partial}
\def\r  {\rho}
\def\ra {\rightarrow}
\def\ri {\right}

\def\s {\sigma}

\def\sq {\sqrt}
\def\t  {\tau}

\def\v {\vec}

\def\> {\rangle}
\def\< {\langle}

\def\hs {\hspace}
\begin{document}

\title{Particle creation by a massless spin-$\f{1}{2}$ field in a warped cosmological braneworld scenario} 

\author{Suman Ghosh\footnote{Electronic address : {\em suman.ghosh@igntu.ac.in}}${}^{}$}
\affiliation{Department of Physics \\
Indira Gandhi National Tribal University, Amarkantak, MP - 484887, India}

%\author{Sayan Kar\footnote{Electronic address : {\em sayan@cts.iitkgp.ernet.in}}${}^{}$}
%\affiliation{Department of Physics and Centre for Theoretical Studies
%\\ Indian Institute of Technology Kharagpur, WB - 721 302, India}

\begin{abstract}
Energy momentum tensor of a quantized massless bulk spin-$\f{1}{2}$ field in five dimensional warped cosmological spacetimes is studied. The four dimensional part of our model represents a warped cosmological thick brane and the scale of the extra dimension is time-dependent. We use a simple ansatz to solve the Dirac equation in the bulk that helps us to compare our results with the known four dimensional case.
Renormalisation of the components of the energy momentum tensor is achieved using adiabatic regularization method. We compute the leading order finite contribution to the stress-energy tensor which is of adiabatic order six. The resulting energy and pressure densities explicitly show the effects of the so-called warping factor and the dynamic extra dimension on the created matter. We show how the energy density produced are localized to form 3-branes along the extra dimension. % 

%The warp factor is chosen to be that of the Randall--Sundrum model. With particular choices for the functional form of the scale factor (and also the function characterising the time evolution of the extra dimension) we obtain the ${\vert \beta_{kq}\vert}^2$, the particle number and energy densities after solving (wherever possible, analytically but, otherwise, numerically) the conformal scalar field equations. The behaviour of these quantities for the massless and massive Kaluza--Klein modes are examined. Our results show the effect of a warped extra dimension on particle creation and illustrate how the nature of particle production on the brane depends on the nature of warping, type of cosmological evolution as well as the temporal evolution of the extra dimension.    
\end{abstract}

%----------------------------------------------------------------------%
%\pacs{04.62.+v, 04.50.-h, 11.10.Kk}

\maketitle

%---------------------------------------------------------------------%

\section{Introduction}

Various classical and quantum aspects of cosmological models with extra spatial dimensions have been analysed extensively since Kaluza-Klein \cite{Kaluza:1921tu,Klein:1926tv}. Initially extra dimensions were introduced as an interesting idea which offers many new possibilities. However, in string theory \cite{G-S-W}, particularly, extra dimensions appear as an essential feature and not just an idea. Recently, the string inspired braneworld models \cite{Rubakov:1983bb,Gogberashvili:1998vx,Gogberashvili:1998iu,Randall:1999ee, Randall:1999vf, ArkaniHamed:1999hk} where our world is viewed as a four dimensional hypersurface (a 3-brane) embedded in five dimensions have been investigated in great detail. Their potential in proposing achievable experimental and observational signatures of extra dimensions is remarkable.  
The warped braneworld model assumes a non-factorisable higher dimensional spacetime geometry and the line element on the 3-brane is {\em scaled} by a so-called {\em warp factor}.
Brane-world models provided a viable resolution of the long-standing hierarchy problem in high energy physics to begin with. 
Later works showed that presence of extra dimensions may also resolve the horizon problem and explain the dark matter and dark energy effects \cite{Brax:2004xh,Maartens:2010ar}. 
Work has also been done to put constraint on the size of the extra dimensions through various experiments and observations, such as through search for deviations from Newton's inverse square law of gravity at short distances \cite{Lee:2020zjt}. %where the fits are in fact done for Yukawa corrections and not for power law corrections which are the ones which arise in braneworld models.
Apart from various classical features, these models also posses rich quantum behaviour. Constraints may also be
found from study of various quantum aspects e.g. particle creation in such backgrounds.

Particle production by quantum fields is an essential quantum consequence of a dynamic background. Analysis of quantum fields in a higher dimensional spacetime with a Kaluza--Klein-like extra dimension and gravitational particle production by scalar fields in braneworld cosmology has been reported in \cite{Bajc:1999mh, RandjbarDaemi:2000cr,Saharian:2005xf,Garriga:1989jx,Nojiri:2003vk,Mak:1999tf,Huang:1989zy,Bambi:2007nz}. 
%Gravitational particle production in braneworld cosmology and its implications have been studied in \cite{Bambi:2007nz}. 
However, investigations on quantum fields are mostly carried out in time-independent braneworld backgrounds where the four dimesional part is flat. Such models do not take into account the cosmological expansion of the 3-brane. Further the scale of the extra dimension could naturally be dynamic. Interestingly, large class of such warped and time-dependent bulk solutions of five dimensional Einstein equations, in presence of various bulk matter fields, were found e.g. in \cite{Ghosh:2008vc}. How a quantised field behave in such background is therefore an important question. 
Further, the amount of matter created may put constraints on various models regarding the size as well as the dynamic nature of the extra dimension.

Quantum field theory in curved spacetime \cite{PT,MW,Wald-QFTCS,Buchbinder,Fulling,BD} aims to study such particle creation process. It lead to discovery of black hole radiation \cite{Hawking:1974sw} and also explains the inhomogeneities in the cosmic microwave background and the origin of the large-scale structure of the Universe \cite{Liddle:2000cg}. The so-called adiabatic vacuum in curved space gives us a notion of particles  that comes closest to the definition of field quanta in Minkowski spacetime. However, particles are defined globally in terms of field modes and thus depends on the large scale structure of spacetime. On the other hand, physical detectors are quasi-local in nature. So, one investigate the stress tensor or the energy momentum tensor (EMT) which is defined locally. 
One important feature of the vacuum expectation value of the EMT components of a quantum field is the presence of the quadratic and logarithmic ultra-violet (UV) divergences in addition to the expected quartic divergence (the Minkowski vacuum energy \cite{SW}).
Various renormalisation methods have been developed to resolve these infinities. The one that is particularly useful for FLRW spacetimes is adiabatic regularization \cite{parker:2012,Parker:1974qw,Zeldovich:1971mw}. In this approach, finite expressions are obtained from the formal infinite ones by subtracting mode by mode (under the integral sign) the diverging terms from the adiabatic expansion of the integrands.

In  \cite{Ghosh:2008zs}, we have studied how the presence of a thick brane \cite{Dzhunushaliev:2009va} with a warped and dynamic extra dimension effect the particle creation rate of a massless bulk quantum scalar field. Further, in \cite{Ghosh:2019ulo}, the author has derived the renormalized energy-momentum tensor (REMT) components for a massless scalar field in such warped cosmological background using adiabatic regularization method. Apart from the dependence on the cosmological expansion, our results showed how the created energy and momentum density is distributed along the extra dimension. %Particularly, for growing warp factor, matter is found to be localised at the location of the thick brane.
Proceeding further, an analysis of a quantised spin $1/2$ field and the corresponding REMT in the context of warped cosmological thick braneworld with a dynamic extra dimension is our objective here. Few studies on fermionic fields in higher dimensions including warped braneworld background has been reported in the literature \cite{Zumino:1983rz, Zhang:1996ee, Hosomichi:1997if, Grossman:1999ra, Brustein:2000hi, Neronov:2001qv, Ichinose:2002kg, Koley:2004at, Germani:2004jf, Macesanu:2005jx, Slatyer:2006un, Li:2009zzs, Barbosa-Cendejas:2015qaa, Smolyakov:2016cvd, Mendes:2017hmv}. However, in case of warped braneworld models, the background geometries were assumed to be static. %i.e. cosmological expansion were not taken into account. 

Recently renormalisation of EMT of spin 1/2 field in a FLRW background is achieved within the framework of adiabatic regularization  \cite{Landete:2013axa,Landete:2013lpa,delRio:2014cha,Ghosh:2015mva,Ghosh:2016epo}. Here, we extend that formalism to analyse a spin $1/2$ field in a five dimensional warped cosmological Universe.
The plan of the article is as follows. In Section II, we write down the Dirac field equations in the context of a FLRW spacetime with an warped extra dimension. %and the adiabatic regularization method to find the REMT developed in \cite{Ghosh:2015mva,Ghosh:2016epo}. 
We assume a particularly simple ansatz for the Dirac field which helps us to compare our results with the four dimensional results. We write the field equations to determine in terms of, which may be called, Zeldovich-Starobinsky (Z-S) variables that were introduced in \cite{Zeldovich:1971mw}. The equations for these Z-S variables can be solved as infinite adiabatic series where each term in the series can be found recursively in a straightforward way.
After solving the field equations, In Section III, we renormalise the EMT components using the adiabatic subtraction method. The leading order contribution to the renormalised energy density is explicitly derived. We emphasize on %how the components of REMT depend on metric functions to figure out 
the distinguishing roles of the warping factor and the time-dependent cosmological scale factors. Apart from the adiabatic regularization we discuss how this created matter density is distributed along the extra dimension, thus providing a notion of localisation of matter creating a {\em physical} thick brane as such.
Finally, in Section IV, we conclude with comments and future plans.

%Higgs as a scalar field.

\section{Dirac field in an warped cosmological background}  

Let us consider the background line element to be generically of the form \cite{Ghosh:2008vc}
\begin{equation} 
ds^2 = e^{2f(y)} a^2(t)[- dt^2 + d\vec{x}^2] + b^2(t) dy^2 ,\label{eq:metric}
\end{equation}
where $ e^{2f(y)}$ is the warp factor, $a(t)$ and $b(t)$ are the scale factors associated with the usual four dimensional spacetime (${\vec x},t$) and the extra dimension ($y$) respectively. $t$ denotes the conformal time.
 
The Lagrangian density of a Dirac field $\Psi(t,\vec{x},y)$ with mass $m$ in a curved background geometry is given by \cite{BD, SW}
\be
{\cal L} = \bar{\Psi} (\G^A \na_A + m)\Psi  , ~~~~ \mbox{where,}~~\bar{\Psi} = i\,\Psi^\dagger \G^0, ~~~~ A = 0,1,2,3,4. \la{eq:lag}
\ee
In curved spacetime, $\G^A$ satisfies the Clifford algebra : $\{\G^A,\G^B\} = g^{AB}$. Note that, $g^{AB} = e^A_a e^B_b \eta^{ab}$, $\G^A = e^A_a \g^a$ and $\{\g^a,\g^b\} = \eta^{ab}$ where $e^a_A$ are the vierbeins, $\g^a$'s are Dirac matrices in flat spacetime and $\eta^{ab}$ is the 5D Minkowski metric $(-1,1,1,1,1)$. The large Latin alphabets denote 5D coordinate indices and the small Latin alphabets denote 5D frame indices.
The Dirac matrices in the Dirac-Pauli representation are given by
\ba
\g^0 = -i \le( \begin{array}{cc} 
I & 0 \\
0 & -I \\ 
\end{array} \ri), ~~~~~ 
\g^i = -i\le( \begin{array}{cc} 
0 & \s^i   \\
-\s^i & 0 \\ 
\end{array} \ri), ~~~~~
\g^y = \le( \begin{array}{cc} 
0 & 1  \\
1 & 0 \\ 
\end{array} \ri)
\ea
where $\s^i$ are the Pauli matrices,
\ba
\s^1 = \le( \begin{array}{cc} 
0 & 1   \\
1 & 0 \\ 
\end{array} \ri),~~\s^2 = \le( \begin{array}{cc} 
0 & -i   \\
i & 0 \\ 
\end{array} \ri),~~\s^3 = \le( \begin{array}{cc} 
1 & 0   \\
0 & -1 \\ 
\end{array} \ri).
\ea
Finally, $\nabla_A = \pa_A - \o_A$ is the covariant derivative. The spin connections $\o_A$ are defined as 
\be 
\o_A = \f{1}{4} e^I_{K}(\pa_A e^{JK} + \G^K_{AB} e^{JB}) \G_I \G_J	\la{eq:spincon}
\ee
where $\G^I_{AB}$ are affine connections (see Appendix \ref{ap:geo}) derived from metric (\ref{eq:metric}).
%%%%%%%%%
%
%The conjugate momenta for Dirac field is given bysee
%\be
%\pi = \f{\pa \sqrt{-g}{\cal L}}{\pa \dot \Psi} = i\sq{-g} \Psi^{\da}
%\ee
%and the corresponding symmeric and real energy momentum tensor is given by
%\be
%T_{AB} = \f{1}{2} [\bar{\Psi} \G_{(A}\na_{B)} \Psi + h.c. ]  \la{eq:emt}
%\ee
The Dirac equation in generic curved spacetime for field $\Psi$ with mass $m$ is given by,
\be
(e^A_a\g^a \nabla_A + m)\Psi = 0. \la{eq:Dirac1}
\ee
%%%%%%%%%%%%%
For metric (\ref{eq:metric}), equation (\ref{eq:Dirac1}) leads to
\be
\le[\g^0 \le(\pa_0 + \f{3\dot{a}}{2a} + \f{\dot{b}}{2b}\ri) + \g^i \pa_i + \g^y \f{a}{b} e^{2f} (\pa_y + 2f') + ma\ri] \Psi = 0 \la{eq:Dirac2}
\ee	
where an over-dot $(\dot{})$ represent derivative wrt $t$ and a prime $(')$ denotes derivative wrt $y$. In the following we will consider massless bulk fields which allow us to proceed with simple separation of variable approach. In the limit $b(t)=1$ and $f(y)=0$, equation (\ref{eq:Dirac2}) matches the 4D case. In the following we solve the field equation for a massless field.

\subsection{Solving the Dirac equation}

To solve equation (\ref{eq:Dirac2}), mode expansion of $\Psi$ is written using annihilation operator for particles ($B_{\v{k}q\l}$) and creation operator for antiparticles ($D^{\da}_{\v{k}q\l}$) as
\be
\Psi(\v{x},y,t) = \sum_{\l=\pm 1} \int d^4k\, \big(B_{\v{k}q\l}U_{\v{k}q\l} + D^{\da}_{\v{k}q\l} V_{\v{k}q\l}\big) ,\la{eq:psi-gen}
\ee
%Let us use the following ansatz \cite{Landete:2013axa}, for  $\Psi$ 
where, $U_{\vec{k}q\l}(t,\v{x},y)$ and $V_{\v{k}q\l}(t,\v{x},y)$ are eigenfunctions or mode solutions ($V_{\vec{k}q\l}$ is obtained by charge conjugation, $V = -\g^2 U^*$, operation on $U_{\v{k}q\l}$) and $\int d^4k$ is a measure to sum over all modes ($\v{k},q$). Following ansatz for mode solutions are chosen for simplicity, in terms of two component spinors, as %\cite{Landete:2013axa}
\ba
U_{\vec{k}q\l}(\v{x},y,t) &=& \f{e^{i \v{k}\cdot \v{x}} G_q(y)}{N} \le( \begin{array}{l} 
h^{I}_{kq}(t) \xi_\l(\v{k})  \\
h^{II}_{kq}(t) \f{\v{\s}\cdot\v{k}}{k} \xi_\l(\v k) \\ 
\end{array} \ri)  \la{eq:U_ansatz} \\
V_{\v{k}q\l}(\v{x},y,t) &=& \f{e^{-i \v{k}\cdot\v{x}} G_q^*(y)}{N} \le( \begin{array}{l} 
-h^{II*}_{kq}(t)  \xi_{-\l}(\v k)  \\
-h^{I*}_{kq}(t) \f{\v{\s}\cdot\v{k}}{k} \xi_{-\l}(\v k) \\ 
\end{array} \ri)  \la{eq:V_ansatz}
\ea 
where $N =  e^{3f/2}(2\pi a)^{3/2} b^{1/2}$ and $\xi_\l(\v k)$ is the normalised two component spinor satisfying $\xi_\l^\da \xi_\l(\v{k}) = 1$, $\f{\v{\s}\cdot\v{k}}{k}\xi_\l(\v{k}) = \l \xi_\l$ and $-i \s^2 \xi^*_\l(\v{k}) = \l \xi^{\da}_{-\l}(\v{k})$ where $\l = \pm 1$ represents the usual four dimensional helicity\footnote{Using either value of helicity or either of $u_{\v{k}\l}$ and $v_{\v{k}\l}$, leads to same end results.}. Our simplistic ansatz will be useful to compare our results with the recently analysed four dimensional counterpart \cite{Landete:2013axa}-\cite{Ghosh:2016epo}. As we will see later, our ansatz would imply a vanishing pressure density along the extra dimension. In order to have chiral fermions as projection of bulk spin $1/2$ field on 3-brane via dimensional reduction, on may choose to work with an ansatz suitable in chiral representation. Let us now proceed to extend our formalism \cite{Ghosh:2015mva,Ghosh:2016epo} to five dimensional bulk spinor fields in a dynamic background. 
The Dirac product is defined as
\be
(u,v) = \int d^4x \sq{h}~ u^\da v
\ee
where, $\sq{h}~ d^4x$ is the volume element on a four dimensional spacelike Cauchy surface.
Normalization conditions satisfied by the mode functions are $(U_{\v{k}q\l}, U_{\v{k'}q'\l'}) = (V_{\v{k}q\l}, V_{\v{k'}q'\l'}) = \d_{\l\l'}\d(\v{k}-\v{k'})\d(q-q')$ and $(U_{\v{k}q\l}, V_{\v{k'},q',\l'}) =(U^{\da}_{\v{k}q\l}, V_{-\v{k'},-q',\l'}) = 0$ which further implies
\be
|h^{I}_{kq}(t)|^2 + |h^{II}_{kq}(t)|^2 = 1, ~~~~~ \int dy~ G^*_q(y) G_{q'}(y)=\d(q-q'). \la{eq:norm_h}
\ee
equation (\ref{eq:norm_h}) ensures that the standard anti-commutation relations for creation and annihilation operators are satisfied. 

Putting equation (\ref{eq:U_ansatz}) in equation (\ref{eq:Dirac2}) and stetting (where $q$ represents momentum along $y$)
\be
e^{f/2}\le( \f{G_q'}{G_q} + \f{f'}{2}\ri) = i\, q ,\la{eq:G}
\ee
we get the following first order coupled differential equations for field dynamics
\ba 
{\dot h}^{I}_{kq} + i \,\O \,h^{II}_{kq} &=& 0,  \la{eq:h1h2}\\
{\dot h}^{II}_{kq} + i\, \O^*\, h^{I}_{kq} &=& 0, \la{eq:h2h1}
\ea
where $\O = k + i \f{a}{b}q $. It is easy to check the Wronskian to be: 
\be
{\dot h}^{I}_{kq}\, h^{II*}_{kq} - h^{I}_{kq}\,{\dot h}^{II*}_{kq} = -i \, \O. \la{eq:wronskian1} 
\ee
Solution of equation (\ref{eq:G}) that satisfies the normalization condition (\ref{eq:norm_h}) is given by,
\be
G_q(y) = \le(\int_{-\infty}^{\infty} e^{-f(y')}\, dy' \ri)^{-1/2} Exp \le[-\f{f}{2} + i\, q \int^y e^{-f(y')/2}\, dy' \ri]. \la{eq:Gsol}
\ee
Thus
\be
|G_q(y)|^2 = \le(\int_{-\infty}^{\infty} e^{-f(y')}\, dy' \ri)^{-1} \, e^{-f(y)} \la{eq:Gsq}
\ee
is independent of `$q$' which is a wave number or mode along $y$. Note that, even for the massless bulk Dirac field, conformal invariance of the field equation (\ref{eq:Dirac2}) is broken for $q \neq 0$ modes and $q/b$ essentially plays the role of `mass' for the on-brane Dirac field as such. 

Proceeding further, equation (\ref{eq:h1h2}) and (\ref{eq:h2h1}) leads to the following decoupled second order equations
\ba
{\ddot h}^I_{kq} - \f{\dot \O}{\O} {\dot h}^I_{kq} + |\O |^2 h^{I}_{kq} &=& 0, \la{eq:h1} \\
{\ddot h}^{II}_{kq} - \f{\dot \O^*}{\O^*} {\dot h}^{II}_{kq} + |\O |^2 h^{II}_{kq} &=& 0. \la{eq:h2}
\ea
%where $\O(t) = \sq{m^2a^2 + k^2}$ and $Q(t) = m \dot{a}$. %Following methodology is applicable to more general backgrounds where equations of similar structure as equations (\ref{eq:h1}) and (\ref{eq:h2}) appear.
The adiabatic vacuum i.e. state of adiabatic order\footnote{A term of $n^{th}$ adiabatic order contains $n^{th}$ derivative of $\O$.} zero (that also satisfies the Wronskian condition) is given by the WKB solution of these field equations (see Appendix \ref{ap:wkb}) as,
\ba
h^{I(0)}_{kq}(t)  =  f\, e_-,~~~~ h^{II(0)}_{kq}(t)  =  f^*\, e_- \la{eq:g1g2}
\ea
with
\be
f = \f{1}{\sq{2}}\le(\f{\O^{~}}{\O^*} \ri)^{\f{1}{4}},~~~~ e_{\pm} = e^{\pm i\int |\O|\, dt}. \la{eq:f1f2e}
\ee
This implies the exact solution can be written as 
\ba
h^{II}_{kq}(t) &=& \a_{kq}(t) h^{I(0)}_{kq}(t) - \b_{kq}(t) h^{II(0)*}_{kq}(t), \la{eq:h1ansatz}\\
h^{II}_{kq}(t) &=& \a_{kq}(t) h^{II(0)}_{kq}(t) + \b_{kq}(t) h^{I(0)*}_{kq}(t) \la{eq:h2ansatz}
\ea
where $\a_{kq}(t)$ and $\b_{kq}(t)$ are the Bogolubov coefficients. The normalization condition (\ref{eq:norm_h}) leads to
\be
|\a_{kq}(t)|^2 + |\b_{kq}(t)|^2 = 1. \la{eq:norm_ab}
\ee
equations. (\ref{eq:h1ansatz}) and (\ref{eq:h2ansatz}) implies, the initial conditions are $\a_{kq}(t_0)=1$ and $\b_{kq}(t_0)=0$. Here, $\a_{kq}(t)$ and $\b_{kq}(t)$ carries the effect of {\em non-adiabaticity}. 
Here, the adiabatic matching \cite{BD} is used to obtain a zeroth order adiabatic vacuum state at a particular initial time, say $t = t_0$. This matching fixes the initial values of the mode functions and their first derivatives at $t_0$. %Note that, adiabatic matching at different times results in different states for the field. 
If one chooses a state of different adiabatic order, for adiabatic matching with the exact state, the associated Bogolubov coefficients (and their initial values at $t_0$) will be different. 
In general, one can write \cite{Landete:2013axa} 
\ba
h^{II}_{kq}(t) &=& \a_{kq}^{(n)}(t) h^{I(n)}_{kq}(t) - \b_{kq}^{(n)}(t) h^{II(n)*}_{kq}(t), \la{eq:h1nansatz}\\
h^{II}_{kq}(t) &=& \a_{kq}^{(n)}(t) h^{II(n)}_{kq}(t) + \b_{kq}^{(n)}(t) h^{I(n)*}_{kq}(t) \la{eq:h2nansatz}
\ea
where $h^{I(n)}_{kq}$ and $h^{II(n)}_{kq}$ are solutions of $n^{th}$ adiabatic order and $\a_{kq}^{(n)}$ and $\b_{kq}^{(n)}$ carries terms of adiabatic order higher than `$n$' i.e. they are time-independent upto adiabatic order `$n$'. Here, we consider the matching with zeroth adiabatic order only. 
Our ansatz, equations. (\ref{eq:h1ansatz}) and (\ref{eq:h2ansatz}) further implies 
\ba
\a_{kq}(t) &=& \Big(f^*\, h^{I}_{kq} + f\, h^{II}_{kq}\Big)e_+,  \la{eq:alpha1} \\
\b_{kq}(t) &=& \Big(f\, h^{II}_{kq} - f^*\, h^{I}_{kq}\Big)e_-. \la{eq:beta1}
\ea 
The number density of spin $1/2$ particles created with momentum ($\vec{k},q$) is then given by,
\be
\langle N_{\vec{k}q}\rangle = \langle B^{\da}_{\vec{k}q\l} B_{\vec{k}q\l}\rangle = \langle D^{\da}_{\vec{k}q\l} D_{\vec{k}q\l}\rangle = |\b_{kq}(t)|^2. \la{eq:no.den.}
\ee
Note that, if one can derive $h^{I}_{kq}$ and $h^{II}_{kq}$ analytically from the field equations, then $|\b_{kq}|^2$ directly follows from equation (\ref{eq:beta1}). However, that would be difficult to find in most cases and there we may use the general formalism presented below.
%
%Note that, the WKB solutions (\ref{eq:g1g2}) obey the following Wronskian condition
%\be
%{\dot h}^{I(0)}_{kq}\, h^{II(0)*}_{kq} - h^{I(0)}_{kq}\,{\dot h}^{II(0)*}_{kq} = F - i k, ~~~~F = \f{k\,Q}{2\O^2}. \la{eq:wronskian2} 
%\ee

Putting equations (\ref{eq:h1ansatz}) and (\ref{eq:h2ansatz}) in equations (\ref{eq:h1h2}) and (\ref{eq:h2h1}), a system of coupled linear first order differential equations is obtained for $\a_{kq}(t)$ and $\b_{kq}(t)$:
\be
\dot{\a}_{kq} = -i F(t)\, \b_{kq}\, e_+^2, ~~~~~~\dot{\b}_{kq} = -i F(t)\, \a_{kq}\, e_-^2, ~~~~~~ \mbox{where} ~~~ F(t) = \f{k\,q \, \dot{c}}{2 |\O|^2}  \la{eq:alphabeta}
\ee
where $c(t) = a(t)/b(t)$. equation (\ref{eq:alphabeta}) implies the absence of $q=0$ modes which is a similar phenomenon as absence of massless particles in a four dimensional conformally flat background. Note that, for bulk scalar fields however, the conformal invariance of the field equation is always broken in presence of a dynamic extra dimension and $q=0$ modes are produced \cite{Ghosh:2008zs}.
The function $F(t)$ is of adiabatic order one and contains $q$ that breaks the conformal invariance of the field equations. 
%In the adiabatic limit ($F\ra 0$), equation (\ref{eq:g1g2}) do satisfy the exact Wronskian 
%(\ref{eq:wronskian1}) and thus represents the adiabatic vacuum.
%equation (\ref{eq:wronskian2}) also implies that the WKB ansatz is not an exact solution (due to presence of $F(t)$) of the field equation during the (non-adiabatic) expansion which is the desired condition for particle creation \cite{parker:2012}. 
Thus $F(t)$ is a measure of non-adiabaticity of the cosmological evolution and plays a key role in determining the amount of matter created in $q\neq 0$ modes.
In equation (\ref{eq:psi-gen}), the creation and annihilation operators carry this non-adiabaticity and so as the Bogolubov coefficients in equations (\ref{eq:h1ansatz} - \ref{eq:h2ansatz}). Thus it is natural for $|\b_{kq}|^2$ to depend on $F(t)$ as particle creation is essentially a quantum consequence of this non-adiabaticity.
%Further, fermions at rest will also not be produced. Similar results were found in \cite{Khanal:2013cjp} using Newman-Penrose formalism. 
%%%%%%%%%%%%%%%%%%%%%%%%%%%%%%%%%%%%%
\subsection{Z-S variables}

To determine $|\b_{kq}|^2$ and the resulting EMT, we follow the formalism developed in \cite{Ghosh:2015mva,Ghosh:2016epo} which is an extension of the methodology introduced by Zeldovich and Starobinskii in \cite{Zeldovich:1971mw}. Define three real and independent variables $s_{kq}$,  $u_{kq}$ and $\t_{kq}$ (or the Z-S variables so to speak), in terms of the two complex variables $\a_{kq}$ and $\b_{kq}$ as,
\ba
s_{kq} = |\b_{kq} |^2,~~~ u_{kq} = \a_{kq}\,\b_{kq}^*\,e_-^2 + \a_{kq}^*\,\b_{kq}\,e_+^2,~~~ %\nn \\
\t_{kq} = i(\a_{kq}\,\b_{kq}^*\,e_-^2 - \a_{kq}^*\,\b_{kq}\,e_+^2). \la{eq:def_sut}  
\ea
For the Z-S variables, we get a system of three linear first order coupled differential equations from equations (\ref{eq:alphabeta}), given by
\ba
\dot{s}_{kq} &=& - F\, \t_{kq}, \la{eq:s}\\
\dot{u}_{kq} &=& -2\,|\O | \, \t_{kq},  \la{eq:u} \\
\dot{\t}_{kq} &=& - 2\,F(1-2\,s_{kq}) + 2\, |\O | \, u_{kq},  \la{eq:t}
\ea
with initial conditions $s_{kq} = u_{kq} = \t_{kq} = 0$ at suitably chosen time $t= t_0$ as discussed earlier. %(e.g.  for de Sitter expansion $t_0 \ra -\infty$). 
The vacuum expectation values of EMT components can be easily written in terms of the Z-S variables which makes the equations (\ref{eq:s}-\ref{eq:t}) as key equations to solve in this formalism.
If we compare equations (\ref{eq:s}-\ref{eq:t}) with their four dimensional counterpart (Appendix \ref{ap:4D}), we see that, behaviour of `$u$' and `$\t$' is exchanged (modulo an overall sign).
%Note that, if one can derive $h^{I}_{kq}$ and $h^{II}_{kq}$ analytically from the field equations, then $|\b_{kq}|^2$ directly follows from equation (\ref{eq:beta1}). Alternatively, one can solve the set of equations (\ref{eq:s}-\ref{eq:t}) numerically (or analytically whenever possible). 

%{\em Renormalisation of stress tensor.~}
To solve the system of equations (\ref{eq:s}-\ref{eq:t}), we assume that the Z-S variables can be expanded in asymptotic series in powers of $|\O|^{-1}$ in the large momenta limit ($|\O| \ra \infty$) as
\be
s_{kq} = \sum_{r=0}^{\infty} s_{kq}^{(r)} ,~~ u_{kq} = \sum_{r=0}^{\infty} u_{kq}^{(r)}, ~~ \t_{kq} = \sum_{r=0}^{\infty} \t_{kq}^{(r)}, ~~~~ r = 0,1,2,... \la{eq:series-sut}
\ee
where the superscript $r$ indicates the adiabatic order. This expansion is valid in the quasi-classical region where $|\dot{\O}|<< |\O|^2$. A term of adiabatic order $r$ contains $r^{th}$ time derivative of $|\O|$ or $c(t)$. Therefore, in the adiabatic regime higher order adiabatic terms contribute negligibly. %Also $|\O|$ is of adiabatic order zero and $F(t)$ is of adiabatic order one as such.
Putting equation (\ref{eq:series-sut}) in equations (\ref{eq:s}-\ref{eq:t}) and equating terms of same adiabatic orders, the expansions of $s_{kq}, u_{kq}, \t_{kq}$ are found to be of the form given below,
\ba
s_{kq} &=& s_{kq}^{(2)} + s_{kq}^{(4)} + s_{kq}^{(6)} + ... , \\
u_{kq} &=& u_{kq}^{(1)} + u_{kq}^{(3)} + s_{kq}^{(5)} + ...  , \\
\t_{kq} &=& \t_{kq}^{(2)} + \t_{kq}^{(4)} + \t_{kq}^{(6)} + ...
\ea
where $s_{kq}^{(r)}, u_{kq}^{(r)}$ and $\t_{kq}^{(r)}$ of different adiabatic orders can be derived from the following recursion relations: 
\ba
\t_{kq}^{(r)} &=& \f{\dot{u}_{kq}^{(r-1)}}{2|\O|}, \la{eq:t^r}\\
s_{kq}^{(r)} &=& - \int F\, \t_{kq}^{(r)} dt, \la{eq:s^r}\\
u_{kq}^{(r+1)} &=& \f{-4F s_{kq}^{(r)} + \dot{\t}_{kq}^{(r)}}{2|\O|}, \la{eq:u^r}
\ea
with  $u_{kq}^{(1)} = \f{F}{|\O|}$ to be the non-vanishing term of the lowest adiabatic order. It is straightforward to solve these equations analytically to arbitrary order. Further, as $k \ra \infty$ and if derivatives of $c(t)$ to any $r^{th}$ order is non zero, we have
\be
s_{kq}^{(r)} \sim k^{-(r+2)},~~ u_{kq}^{(r)} \sim k^{-(r+1)}, ~~ \t_{kq}^{(r)} \sim k^{-(r+1)}. \la{eq:lim_sut}
\ee
The leading order terms would imply the well-known logarithmic (for $r=4$) and quadratic (for $r=2$) UV divergences of the total energy and pressure density. %Note that there is no quadratic divergence that appears for scalar fields \cite{PT}. 
%Further, as $k \ra \infty$, the leading term in particle number density $s_{kq}$ decays as $k^{-4}$ irrespective of the functional form of the scale factor. 

%\vspace{.25cm}
\section{The energy-momentum tensor and renormalisation}

%{\em Energy-momentum tensor.~}
The expectation of the energy-momentum tensor operator wrt adiabatic vacuum for the massless Dirac field in curved space is given by
\be
\< T_{AB} \> = \Big{\< } \f{1}{2} \Big[\bar{\Psi}\G_{(A}\na_{B)}\Psi + h.c. \Big] \Big{\> } .\la{eq:stress}
\ee
Note that 
\be
\< \bar{\Psi} \g^a \na_b \Psi \> = \sum_{\l = \pm 1} \int d^4k \, \bar{v}_{kq\l} \g^a \na_b v_{kq\l} .
\ee 
Thus the vacuum expectation values of the the independent and non-vanishing components of EMT operator are derived as
%, the energy density $T^0_0$ and pressure density $T^i_i$ are
%\ba
%T^0_0 &=& -\f{i}{2a} \Big(\bar{\Psi}\g^0\dot{\Psi} - \dot{\bar{\Psi}}\g^0\Psi\Big), \la{eq:T00}\\
%T^i_i &=& \f{i}{2a} \Big(\bar{\Psi}\g^i\Psi' - \bar{\Psi}'\g^i\Psi\Big), \la{eq:Tii}
%\ea
%where ($'$) denotes derivative with respect to $x^i$. Using equations (\ref{eq:psi-gen}-\ref{eq:-psi_ansatz}), vacuum expectation value of the above quantities can be written as
\ba
\< T^0_{~0} \> \equiv -\r &=& -\f{1}{2\pi^2 e^{3f} a^3 b} \, |G_q|^2 \int dk\,k^2 \, \r_{kq}  ,\la{eq:<T00>}\\
\< T^i_{~i} \> \equiv p_i &=& \f{1}{2\pi^2 e^{3f} a^3 b} \, |G_q|^2 \int dk\,k^2 \, p_{kq} , \la{eq:<Tii>}\\
\< T^y_{~y} \> \equiv p_y &=& 0 ,\la{eq:<T55>} \\
\< T^0_{~y} \> = \< T^y_{~0} \> &=& \f{1}{2\pi^2 e^{3f} a^3 b} \, \f{|G_q|^2}{e^{f/2}} \int dk\,k^2 ,\la{eq:<T05>}
\ea
where $\rho_{kq}$ and $p_{kq}$, for each mode, are given respectively as,
\ba
\r_{kq} &=& -\f{i}{a} \Big(h^{I}_{kq} \dot{h}^{I*}_{kq} + h^{II}_{kq} \dot{h}^{II*}_{kq} - h^{I*}_{kq} \dot{h}^{I}_{kq} - h^{II*}_{kq} \dot{h}^{II}_{kq}\Big), \la{eq:r_{kq}} \\
p_{kq} &=&  \f{2k}{3a} \Big(h^{I}_{kq} h^{II*}_{kq} + h^{I*}_{kq} h^{II}_{kq}\Big). \la{eq:p_{kq}}
\ea
Using equations (\ref{eq:h1ansatz}), (\ref{eq:h2ansatz}), (\ref{eq:norm_ab}) and (\ref{eq:def_sut}), we can rewrite equations (\ref{eq:r_{kq}}) and (\ref{eq:p_{kq}}) as
\ba
\r_{kq} &=& \f{2|\O|}{a}\big(1-2s_{kq}\big), \la{eq:rho_{kq}_sut} \\
p_{kq} &=& -\f{2k}{3a} \le[\f{k}{|\O|}(1- 2s_{kq}) - \f{a\,q}{b\,|\O|} \t_{kq}\ri]. \la{eq:p_{kq}_sut}
\ea
The vacuum energy (in absence of any gravitationally created particles when $s_{kq} = u_{kq} = \t_{kq} = 0$) matches with the Minkowski scenario. 
$\< T^\m_{~\m} \> $ components are functions of Z-S variables and contain {\em all} the UV divergences as such. 
Note that, $\< T^y_{~y} \> $ or the pressure density along the extra dimension vanishes identically. As mentioned earlier, this situation may change with a more general ansatz replacing equation \ref{eq:U_ansatz}.
Further $\< T^0_{~y} \> $ diverges quartically and is not dependent on Z-S variables. So the renormalized value of the off-diagonal components are simply zero.
To remove the divergences in $\< T^\m_{~\m} \> $ components, we need to subtract the Minkowski vacuum contribution as well as leading terms upto necessary adiabatic order from the expansion of $s_{kq}$, $u_{kq}$ and $\t_{kq}$. 

%\vspace{.25cm}

Let us briefly discuss the validity of adibatic approximation in the present context. Define the on-brane Hubble parameter to be $h = \dot{c}/{c^2}$. Then, as long as $q/b << h$, the higher order adiabatic corrections (which go as powers of $|\dot\O|/ |\O|^2$, see Appendix \ref{ap:sut}) become smaller than the lower order ones even for  $k = 0$ mode. Thus, the adiabatic expansion is legitimate for all $k$-modes \cite{Kaya:2011yu}.
Note that, with $q=0$, the adiabatic expansion fails for modes with $k < h$ (i.e. for on-brane super-horizon modes). However, the expansion is still valid for the modes satisfying $k >> h$ (i.e. for on-brane sub-horizon modes). Thus the momentum integrals should be performed in the interval ($k_* , \infty$), for some $k_*>>h$. Since the divergences are essentially caused by large $k$-modes, for $q \neq 0$, we expect the adiabatic approximation to be valid even if we set the lower $k$-limit to be zero while evaluating the integrals.  

%\subsection{renormalisation}

We are interested in the distribution of energy and momentum density along $y$ on four dimensional hypersurfaces for specific $q$ or $q/b$ which plays the role of effective 4D mass. Thus we do not attempt to regularize the divergence coming from summing over all $q$.
%Note that the total particle number density of specific mass with summed over momenta is given by
%\be 
%N_m = \f{1}{(2\pi e^f\,a)^3\, b} \int dq \int d^3k \,s_{kq}. % \approx  \f{1}{2\pi^2 a^3} \int dk\,k^2\, s_{kq}^{(4)} 
%\la{eq:totalPND}
%\ee
%and has logarithmic divergence unlike the 4D case where it is finite \cite{}.
To find the projected renormalized energy and pressure density one needs to subtract from $\< T^\m_{~\m} \> $ components the vacuum contribution (which is present even in flat space and in absence of particles) and Z-S variables terms upto fourth adiabatic order. This leads to
%\ba
%{\< T^0_0 \> }_{ren} &=& \f{2}{\pi^2 a^4} \int dk\, k^2\, \O \Big(s_{kq} - s_{kq}^{(2)}\Big), \la{eq:r_{kq}_ren} \\
%%&\approx & \f{2}{\pi^2 a^4} \int dk\, k^2\, \O\, s_{kq}^{(6)}, \la{eq:r_{kq}_ren6} 
%{\< T^i_i \> }_{ren} &=&  \f{1}{3\pi^2 a^4} \int dk \f{k^3}{\O} \Big[-2k \big(s_{kq}  - s_{kq}^{(2)}\big) \nn \\
%&& \hspace{1.5cm} + ma \big(u_{kq}  - u_{kq}^{(2)}\big)\Big]. \la{eq:p_{kq}_ren}
%%&\approx &  \f{1}{3\pi^2 a^4} \int dk \f{k^3}{\O} \Big[2k\, s_{kq}^{(6)} + ma\, u_{kq}^{(6)}\Big] .  \la{eq:p_{kq}_ren6}
%\ea
%equations (\ref{eq:r_{kq}_ren6}) and (\ref{eq:p_{kq}_ren6}), gives the energy and momentum density  respectively, upto the leading order.
%%%%%%%%%%%%%%%%%%%%%
\ba
\r_{{}_{Ren}} &=& \f{2\,|G_q|^2 }{\pi^2 e^{3f} a^4 b} \, \int dk\,k^2 \, |\O| \Big(s_{kq} - s_{kq}^{(2)} - s_{kq}^{(4)}\Big), \la{eq:r_{kq}_ren_gen} \\
&\approx & \f{2\,|G_q|^2 }{\pi^2 e^{3f} a^4 b} \, \int dk\,k^2 \, |\O|\, s_{kq}^{(6)}, \la{eq:r_{kq}_ren6} \\
p_{{i}_{Ren}} &=&  \f{\,|G_q|^2 }{\pi^2 e^{3f} a^4 b} \, \int dk \f{k^3}{|\O|} \Big[2k \big(s_{kq}  - s_{kq}^{(2)} - s_{kq}^{(4)}\big) \nn \\
&& \hspace{3.5cm} + \f{q\,a}{b} \big(\t_{kq}  - \t_{kq}^{(2)} - \t_{kq}^{(4)}\big)\Big]. \la{eq:p_{kq}_ren_gen}\\
&\approx &   \f{\,|G_q|^2 }{\pi^2 e^{3f} a^4 b} \, \int dk \f{k^3}{|\O|} \Big[2k\, s_{kq}^{(6)} + \f{q\,a}{b}\, \t_{kq}^{(6)}\Big] .  \la{eq:p_{kq}_ren6}
\ea
In equations (\ref{eq:r_{kq}_ren6},\ref{eq:p_{kq}_ren6}), we have approximated the energy density by the leading order term which is of sixth adiabatic order. It can be easily shown that $ \int dk\,k^2 \, |\O| \,s_{kq}^{(r)} \propto q^{4-r} $. Thus the energy density of more and more `massive' modes are suppressed exponentially. Below we discuss the implications of the leading term in energy density which is of adiabatic order six and an on-brane conformal anomaly derived in the limit $q\ra0$.

\section{Localisation and time evolution of matter density}

The integrals in equations (\ref{eq:r_{kq}_ren6}, \ref{eq:p_{kq}_ren6}) can be explicitly performed using sixth order adiabatic terms. We find the energy density to be,
\ba
\r_{{}_{Ren}} &=& \f{|G_q|^2\, e^{-3f} }{20160\pi^2 q^2 c^{12} b^5} \le[685\dot{c}^6 - 1800 \dot{c}^4\ddot{c} + 660 c^2 \dot{c}^3\dddot{c} + 750\dot{c}^2\ddot{c}^2  - 144 c^3 \dot{c}^2\ddddot{c} \ri. \nn \\
& & \hs{3cm} \le. + 80c^3 \ddot{c}^3 + 9c^4\dddot{c}^2 - 18c^4 \ddot{c}\ddddot{c} -240c^3\dot{c}\ddot{c}\dddot{c} + 18c^4\dot{c} \f{d^5c}{dt^5} \ri] .\la{eq:rho-result}
\ea
%Note that, in the specific case of FLRW background, it is enough to subtract only upto second adiabatic order. However, in more generic spacetimes the fourth-order adiabatic terms may give rise to proper UV divergences \cite{Christensen:1978yd} and therefore the standard approach of regularization considers the fourth order adiabatic terms as {\em potentially divergent} \cite{PT,BD}.
This shows that the energy density decreases with increasing $q$. The functional dependence of the REMT components on time and extra dimension can be analysed graphically as a two dimensional surface in $T,t,y$ space. However, since the dependence on $t$ and $y$ are separated in equation (\ref{eq:rho-result}), we have an advantage of addressing them separately. A more genral ansatz for the Dirac field could have resulted in a non-separable energy density as is found for the scalar field in  \cite{Ghosh:2019ulo}. Note that, equation (\ref{eq:rho-result}), can be used to constrain $f(y)$ and $b(t)$, given an $a(t)$ (see \cite{Ghosh:2016epo} for instance), as the `on-brane' co-moving energy density has to be less than the critical energy density, i.e $ \r^{brane}_{{}_{Ren}} / \r_{{}_c} \leq 1 $.
Below, we discuss the localisation or distribution of the created matter and their time evolution separately.

\subsection{Localisation of energy density}

In present case, the $y$-dependence in $\r_{{}_{Ren}}$, for example, comes from an over all factor of $|G_q|^2\, e^{-3f}$. equation (\ref{eq:Gsq}), suggests that this overall factor is proportional to $e^{-4f}$. To be specific, below we discuss the role of warp factor on destribution of created matter density along the extra dimension using standard ansatz for the so-called growing and decaying warp factors \cite{Dzhunushaliev:2009va}.  
%In general, $\s$-dependence in $\< T_{AB}\> $ comes from the factors $K_A^2(k_\s;\s)$ (for $A = t,i,\s $) except the overall factor. 
In case of growing warp factor, we set $f(y) = \log(\cosh y/y_0)$ (here $y_0$ defines a characteristic length scale along the extra dimension). Then, equation (\ref{eq:Gsol}) implies
\be
G_q(y) = \le(\pi \, \cosh{\f{y}{y_0}}\ri)^{-1/2} \exp \big( 2q\,y_0 \F[i y/2y_0, 2]\big).
\ee
Where `$\F$' is the elliptic integral of the first kind.
Thus, in this case, we have 
\be
|G_q|^2\,e^{-3f} \sim e^{-4f} = \sech^4 \f{y}{y_0},
\ee
which is maximum at $y = 0$ and decays exponentially with increasing $|y|$. This implies that most of the matter density created in mode $q$ is localised near $y=0$ and can be thought of building a `thick' brane as such.

For the decaying ($f(y) = - \log(\cosh y/y_0) $) warp factor, the solution given by equation (\ref{eq:Gsol}) is not normalisable. However, if we assume the extra dimension to be finite and set $y \leq |y_c|$ we get a normalised solution given by
\be
G_q(y) = \le( \le[\sinh\f{y}{y_0}\ri]^{y_c}_{-y_c}  \sech{\f{y}{y_0}}\ri)^{-1/2} \exp \big(2q\,y_0 \E[i y/2y_0, 2]\big).
\ee
Where `$\E$' is the elliptic integral of the second kind. This type of hard boundary would imply $G_q(y)$ will vanish at and beyond $\pm y_c$ and $G''(\pm y_c)$ will contain $\d(y\pm y_c)$ which must then be accounted for in the Lagrangian by putting $\d(y\pm y_c)$ appropriately. More importantly, the finite size of the extra dimension implies that particles with discrete $q$ values are created with
\be
q = \f{(2n+1)\pi}{4\,y_0 \E[i y_c/2y_0, 2]}, ~~~~ n = 0, \pm 1, \pm 2,...  .
\ee
Then the the overall $y$-dependence for decaying warp fator, given by
\be
\f{|G_q|^2}{e^{3f}} \sim \cosh ^4\f{y}{y_0},
\ee
would grows exponentially with increasing $|y|$. Thus the most amount of matter (of all possible modes) is created near the boundaries $y=\pm y_c$. This scenario could represent two `thin' branes formed near or at the boundary walls along a finite extra dimension.
%We can do this even before the adiabatic subtraction as regularization will only effect Z-S variables which do not depend on $\s$. 

\subsection{Time evolution of energy density}

Let us now turn to the time evolution of the energy density and write 
\be
\r_{{}_{Ren}} = \f{|G_q|^2 \r_t}{e^{3f} q^2} 
\ee
where $\r_t$ carries all the time dependence in equation (\ref{eq:rho-result}). To have an idea of what role the dynamic scale of the extra dimension may play, let us look at three simple cases where the 3-brane represents a asymptotically radiative Universe at $t \ra \pm \infty$ with a bounce at $t=0$.

%\subsubsection{Case-1: $a(t)= \sq{1+(t/t_0)^2}$, $b(t) = 1$}
\begin{itemize}
\item Case-1: $a(t)= \sq{1+(t/t_0)^2}$, $b(t) = 1$
%\end{itemize}
%\noindent Case-1: $a(t)= \sq{1+(t/t_0)^2}$, $b(t) = 1$.\\

Here, the scale of the extra dimension is constant. The energy density profile in this case is similar to a 4D scenario with no extra dimension (modulo the overall factor). Figure (\ref{fig:case1})\footnote{For graphical computation we have set $t_0=1$} shows that, at the bounce, energy density has a local minimum. Further, for the bounce to be followed by a late time radiative brane, the energy density eventually becomes negative and at some $|t|=t_c$, hits the minima and asymptotically reaches zero as $|t| \ra \infty$. Note that, the negative energy density regime is a characteristic feature of the bounce as that region disappears for $a(t) \propto t$. %\footnote{Negative $\r_t$ region is absent for $a(t) \propto t$.}
\begin{figure}%[!ht]
\hspace{1cm}
\includegraphics[width=2.2in]{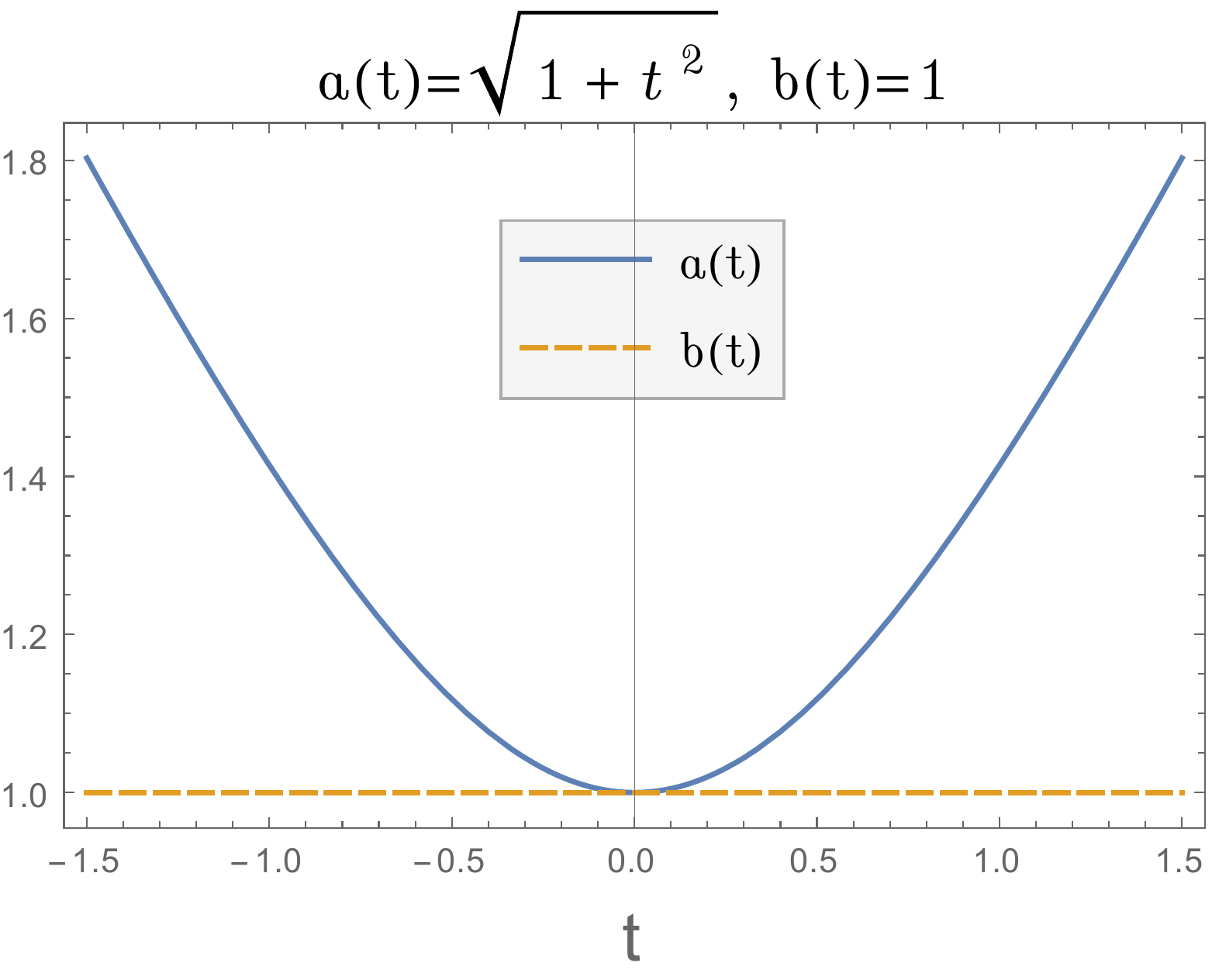} 
\hspace{1cm}
\includegraphics[width=2.5in]{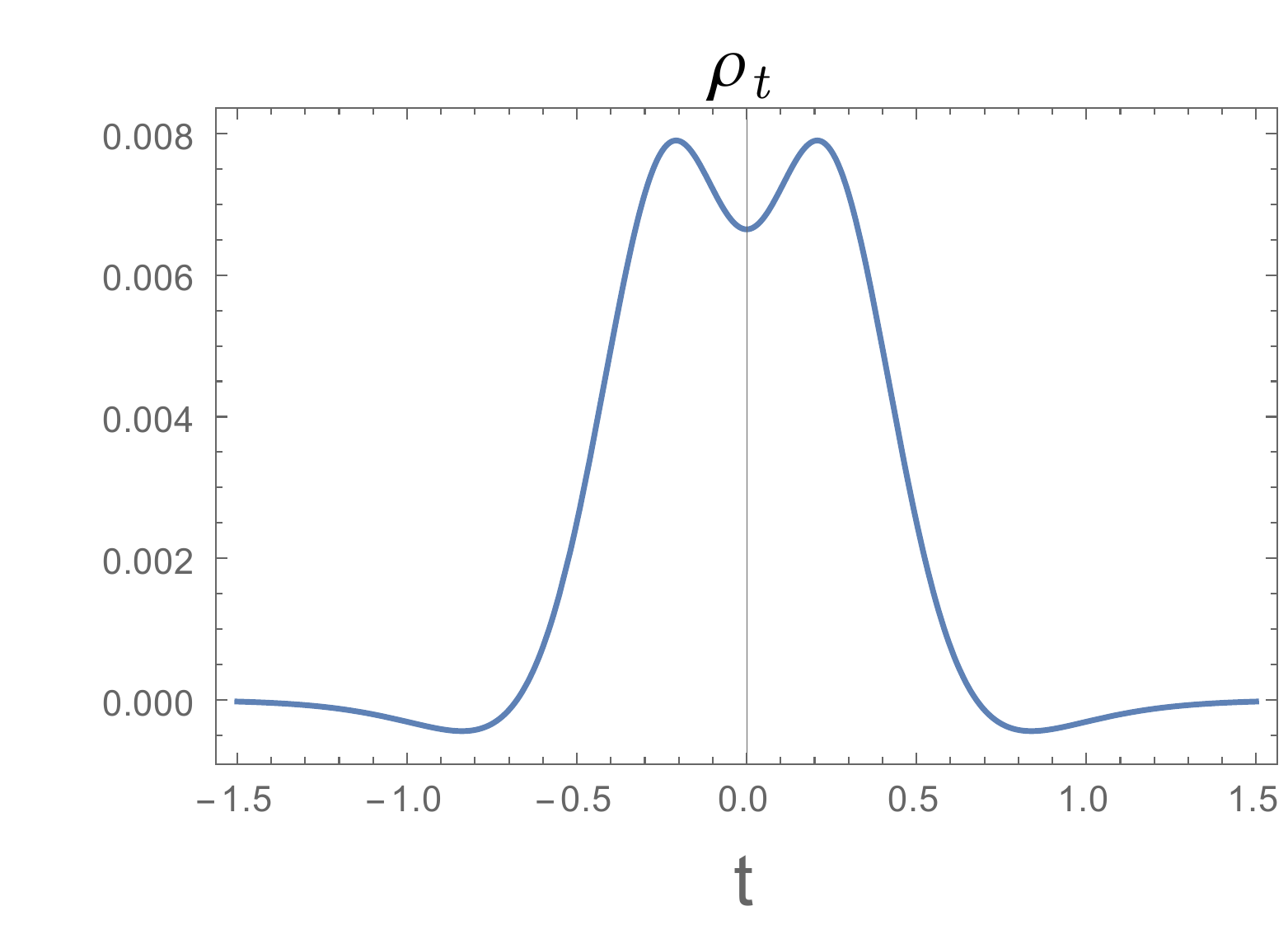} %[width=2.2in,angle=-90] 
\caption{Time dependence of scale factors and the renormalised energy density for Case-1.} \label{fig:case1} 
\end{figure}

%figure (\ref{fig:case1})\footnote{For graphical computation we have set $t_0=1$} shows that for the bounce to be followed by a late time radiative brane, the energy density eventually becomes negative\footnote{Negative $\r_t$ region is absent for $a(t) \propto t$.} and at some $|t|=t_c$,  hits minima and asymptotically reaches zero as $|t| \ra \infty$. 
%%%%%%%%%%%%%%%%%%%%%%%%%%%%%%%%%%%%%%%%%%%%%%%%%%

%\subsubsection{Case-2: $a(t)= \sq{1+(t/t_0)^2}$, $b(t) = \sq{\f{t_0^2+t^2}{4t_0^2+t^2}}$}

%\begin{itemize}
\item Case-2: $a(t)= \sq{1+(t/t_0)^2}$, $b(t) = \sq{\f{t_0^2+t^2}{4t_0^2+t^2}}$
%\end{itemize}
%\noindent Case-2: $a(t)= \sq{1+(t/t_0)^2}$, $b(t) = \sq{\f{t_0^2+t^2}{4t_0^2+t^2}}$.\\

Here the scale of the extra dimension also has a minimum at $t=0$ (with $b(t)|_{t=0}=a(t)|_{t=0}/2$) and grows with $|t|$ and $b(t\ra \pm \infty)$ equals $a(t)|_{t=0}$, i.e. the scale of the extra dimension, although growing, always remains smaller than the cosmological scale factor. Figure (\ref{fig:case2}) shows that presence of such expanding extra dimension flattens the local minima in energy density that occurs at $t=0$ and also lowers the value of negative minima $\r_t(t_c)$. Thus an expanding extra dimension essentially contributes with a positive energy density and may give rise to arbitrarily small negative energy density at late times while the 3-brane still has a bounce.
\begin{figure}%[!ht]
\hspace{1cm}
\includegraphics[width=2.2in]{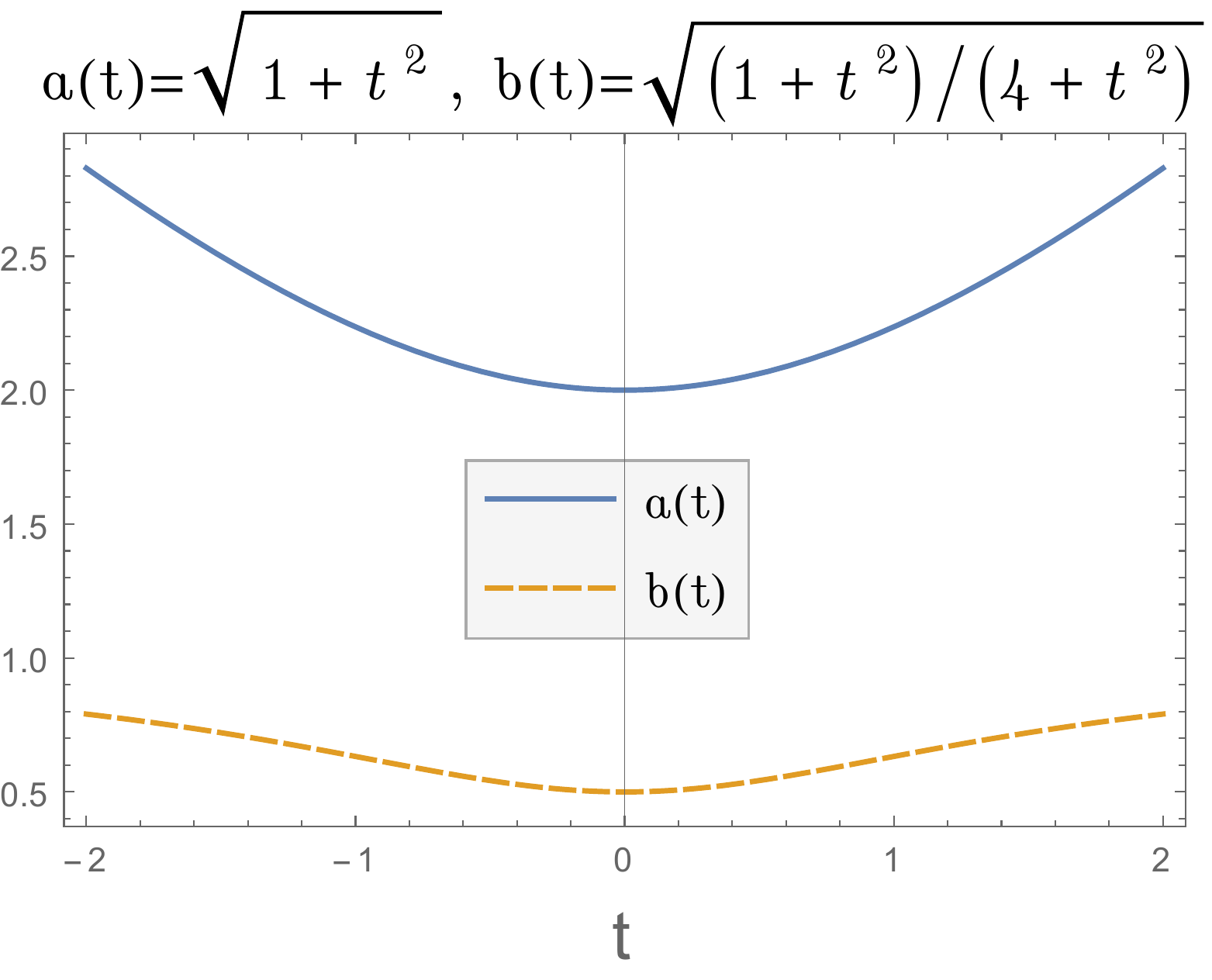} 
\hspace{1cm}
\includegraphics[width=2.5in]{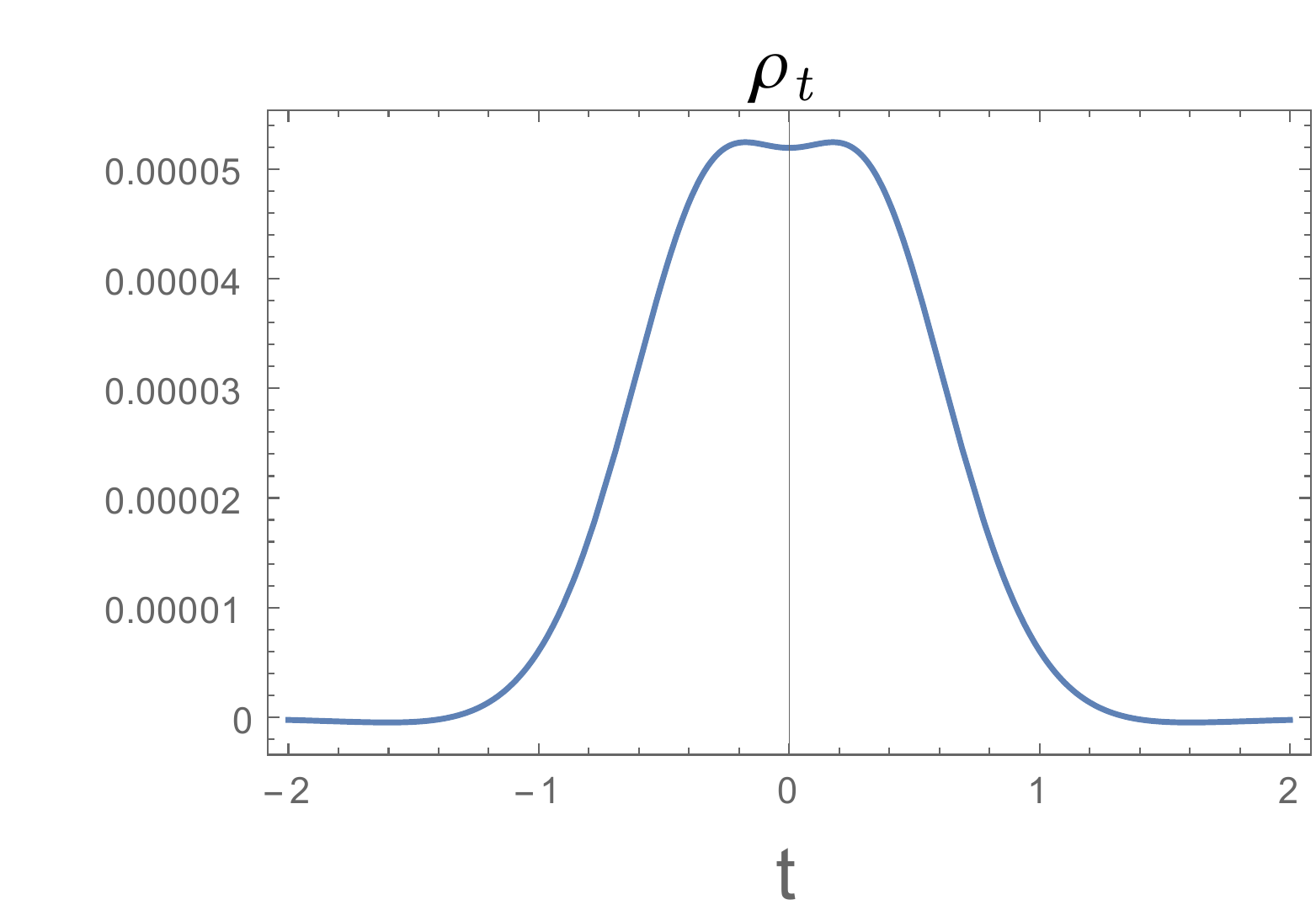} %[width=2.2in,angle=-90] 
\caption{Time dependence of scale factors and the renormalised energy density for Case-2.} \label{fig:case2} 
\end{figure}

%%%%%%%%%%%%%%%%%%%%%%%%%%%%%%%%%%%%%%%%%%%%%%%%%%%%

%\subsubsection{Case-3: $a(t)= \sq{1+(t/t_0)^2}$, $b(t) = \sq{\f{t_0^2+t^2}{t_0^2+4t^2}}$}
%\begin{itemize}
\item Case-3: $a(t)= \sq{1+(t/t_0)^2}$, $b(t) = \sq{\f{t_0^2+t^2}{t_0^2+4t^2}}$
%\noindent Case-3: $a(t)= \sq{1+(t/t_0)^2}$, $b(t) = \sq{\f{t_0^2+t^2}{t_0^2+4t^2}}$

Here the scale of the extra dimension has a maximum at $t=0$ (with $b(t)|_{t=0}=a(t)|_{t=0}$) and shrinks with increasing $|t|$ and $b(t\ra \pm \infty)$ equals $a(t)|_{t=0}/2$. Here also, the scale of the extra dimension always remains smaller than the cosmological scale factor. Figure (\ref{fig:case3}) shows that presence of such shrinking extra dimension enhances the local minima in energy density that occurs at $t=0$ and also the value of negative minima $\r_t(t_c)$. It is interesting to note that the analytic solutions found in \cite{Ghosh:2008vc} are mostly of this type.
\begin{figure}%[!ht]
\hspace{1cm}
\includegraphics[width=2.2in]{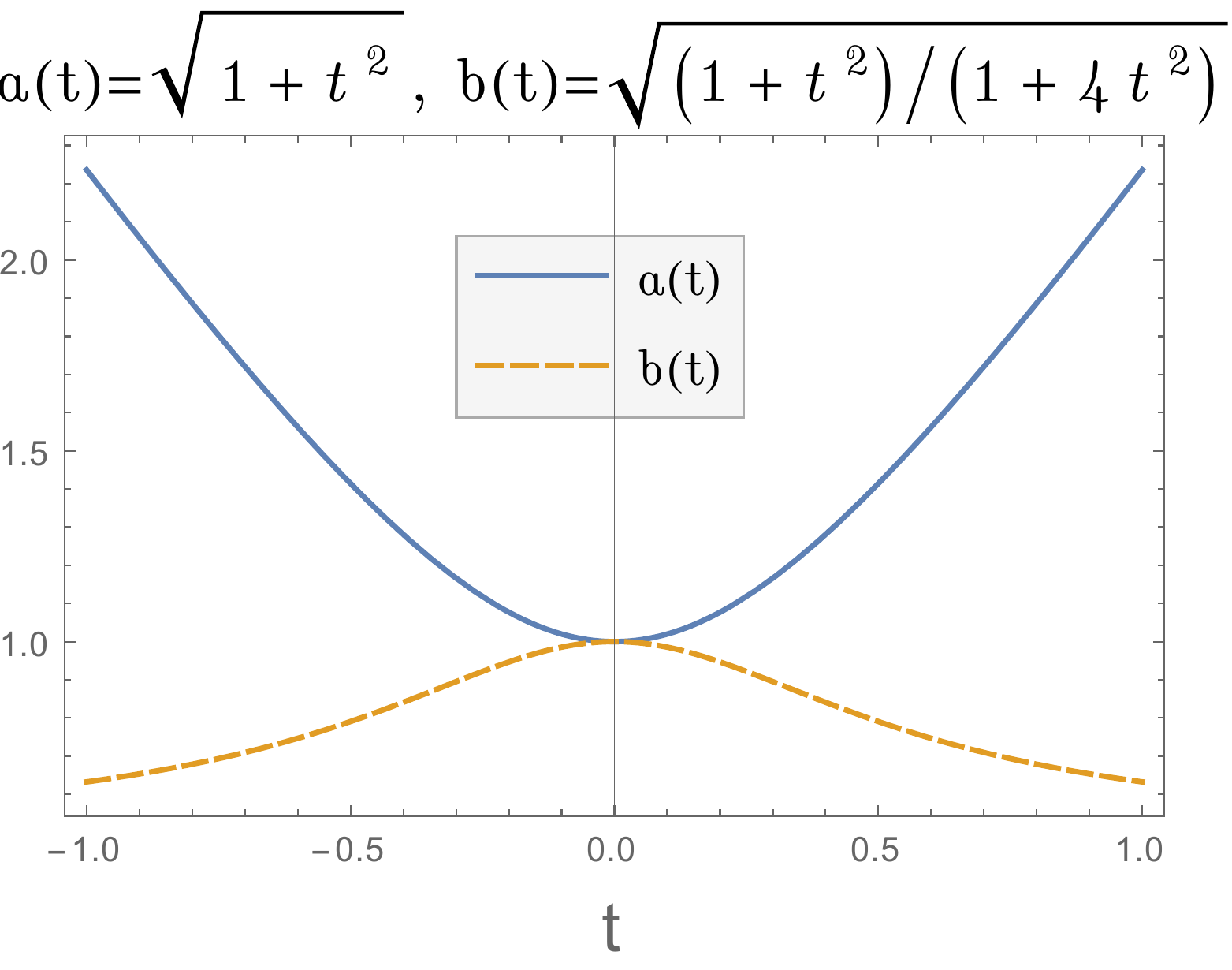} 
\hspace{1cm}
\includegraphics[width=2.5in]{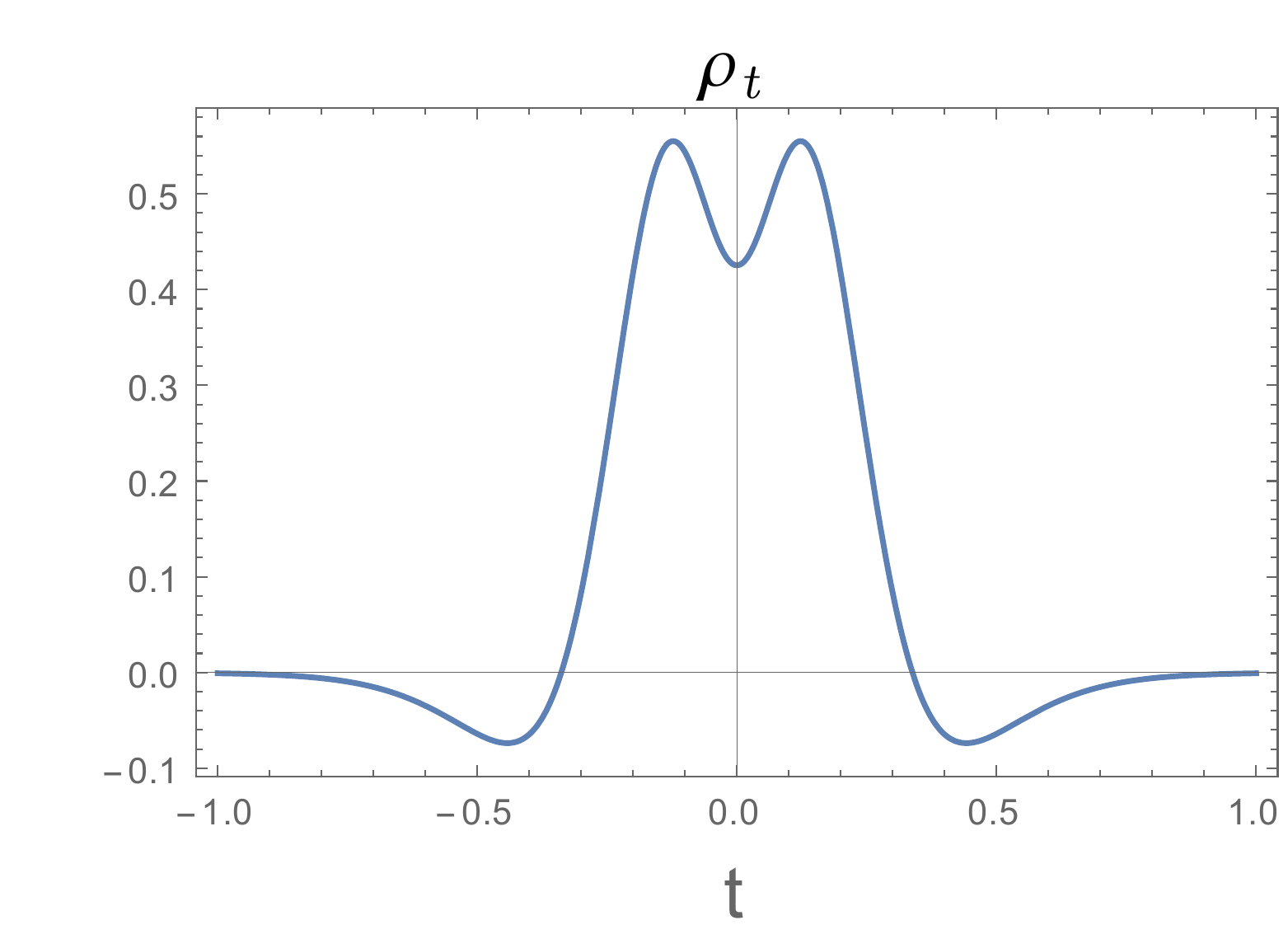} %[width=2.2in,angle=-90] 
\caption{Time dependence of scale factors and the renormalised energy density for Case-3.} \label{fig:case3} 
\end{figure}
\end{itemize}
Thus the dynamic property of the extra dimension does have an impact on the matter creation sourced by a spin-$1/2$ field. Implications when the energy density becomes negative is particularly interesting. Further detailed study on this aspect (including pressure density profiles as well), for various cosmological scenario like de Sitter expansion which is thought to be driven by a cosmological constant with negative energy density, is needed. As mentioned earlier, a more general ansatz for the Dirac field may provide more insights. %It is also suggestive that one should study the pressure density profiles as well in a radiation dominated and de Sitter brane dynamics. 

\section{Conformal anomaly on the brane}
%{\em Conformal and axial anomalies.~}
In an even dimensional conformally flat spacetime, the trace of the energy momentum tensor of a Dirac field with mass $m$ given by $T^\m_{~\m} = m \bar{\Psi} \Psi$ (where $\m$ is summed over), vanishes in the $m \ra 0$ limit.
%\be
%T^\m_\m = m \bar{\Psi} \Psi. \la{eq:Tmm}
%\ee
However, renormalisation procedure renders the quantum counterpart of $T^\m_{~\m}$ finite. This phenomenon is known as the conformal anomaly. 
Similarly, on the brane (which is conformally flat) one can derive an effective conformal anomaly in the limit $q \ra 0$.
The vacuum expectation value of the trace of {\em on-brane} stress tensor in mode $q$ is
\be
\< T^\m_{~\m}  \>  = \f{|G_q|^2 }{2\pi^2 e^{3f} a^3 b} \, \int dk\,k^2 \, \< T\> _{k} \la{eq:<Taa>}
\ee
with
\ba
\< T \> _{k} &=&  2\f{q}{b} \le[\f{a\,q}{b|\O|}(1-2s_{kq}) + \f{k}{|\O|} \t_{kq}\ri] \la{eq:<Tmm>k_sut}
\ea
%using  equations (\ref{eq:h1ansatz}), (\ref{eq:h2ansatz}), (\ref{eq:norm_ab}) and (\ref{eq:def_sut}). Alternatively, equation (\ref{eq:<Tmm>k_sut}) follows from the identity $\< T^\m_\m \> _{kq} = \r_{kq} + 3p_{kq}$ too.
To renormalise this trace, the usual method of subtracting the vacuum contribution and terms upto fourth adiabatic order gives,
\ba
\< T^\m_{~\m}\> _{{}_{Ren}} &=&  \f{|G_q|^2 \, q}{\pi^2 e^{3f} a^3 b^2}  \int dk\,k^2 \, \le[-\f{2a\,q}{b|\O|}\le(s_{kq} - s_{kq}^{(2)} - s_{kq}^{(4)}\ri)\ri. \nn\\ 
&& \hs{4cm}\le. + \f{k}{|\O|} \le(\t_{kq} - \t_{kq}^{(2)} - \t_{kq}^{(4)}\ri)\ri].	
\ea
The above expression in the $q\ra 0$ limit gives,
\ba
\lim_{q\ra 0} \< T^\m_{~\m}\> _{{}_{Ren}} &=& \lim_{q\ra 0} \f{|G_q|^2 \, q}{\pi^2 e^{3f} a^3 b^2}  \int dk\,k^2 \, \le[-\f{2a\,q}{b|\O|} s_{kq}^{(4)} + \f{k}{|\O|} \t_{kq}^{(4)}\ri]~~~~ \\
&=&  \f{|G_q|^2}{e^{3f} b^5}\, \le( \f{-11\dot{c}^4 + 29 c \dot{c}^2\ddot{c} - 12  c^2\dot{c}\dddot{c} - 9 c^2 \ddot{c}^2 + 3 c^3\ddddot{c}}{240\pi^2 c^8}\ri) , \la{eq:<Tmm>ren1}
\ea
where we have used explicit expressions of $s_{kq}^{(r)}$ and $u_{kq}^{(r)}$ given in Appendix \ref{ap:sut}.
%as only the fourth order term in the expansions of $s_{kq}(t)$ and $\t_{kq}(t)$ survives in the $q\ra 0$ limit.	%One can also derive equation (\ref{eq:<Tmm>ren1}) directly from equations (\ref{eq:r_{kq}_ren}) and  (\ref{eq:p_{kq}_ren}). 
%Using explicit expressions of $s_{kq}^{(4)}$ and $u_{kq}^{(4)}$ (Appendix \ref{app-2}) in equation (\ref{eq:<Tmm>ren1}), we get\footnote{Here we correct an overall sign mistake made in \cite{Ghosh:2015mva}.}
%\be
%\< T^\m_\m \> _{Ren} = \f{|G_q|^2}{e^{3f} b^5}\, \le( \f{-11\dot{c}^4 + 29 c \dot{c}^2\ddot{c} - 12  c^2\dot{c}\dddot{c} - 9 c^2 \ddot{c}^2 + 3 c^3\ddddot{c}}{240\pi^2 c^8}\ri). \la{eq:<Tmm>ren2}
%\ee
Note that, equation (\ref{eq:<Tmm>ren1}) represent the trace anomaly of a spin 1/2 field with mass $m\equiv q/b(t)$ (in the $m \ra 0$ or $q \ra 0$ limit) in a four dimensional warped FLRW universe with effective scale factor $c(t)=a(t)/b(t)$. If we set $b(t)=1$ and $f(y)=0$, equation (\ref{eq:<Tmm>ren1}) matches exactly with the four dimensional result \cite{PT,MW,Wald-QFTCS,Buchbinder,Fulling,BD}. However in this case, the effective trace anomaly also has an overall factor which is proportional to $e^{-4f}$ and is different on different location along the extra dimension. Effect of such overall factor in the back reaction problem would be interesting to investigate.

%Note that the conformal anomaly can be expressed in terms of the curvature invariants as \cite{PT,BD},
%\be
%\< T^\m_\m \> _{Ren} = \f{1}{(4\pi)^2} \le(A\, C_{\a\b\g\d}C^{\a\b\g\d} + B\, G + C\, \square R \ri), \la{eq:Tmm_gen}
%\ee
%where $C_{\a\b\g\d}$ is the Weyl tensor, $R$ is the Ricci scalar and $G$ is the Gauss-Bonnet invariant, given by $G = -2(R_{\a\b}R^{\a\b} - R^2/3)$ with $R_{\a\b}$ being the Ricci tensor. For conformally flat spacetimes, Weyl tensor vanishes identically. Equating equation (\ref{eq:Tmm_gen}) with equation (\ref{eq:<Tmm>ren2}), we get $B = 11/360$ and $C=-1/30$, which agrees with the known results \cite{PT,BD}. %This provides a cross-check for the viability of the methodology presented here. 

%Similarly one can check that the axial anomaly vanishes as expected \cite{Ghosh:2015mva}.

%\section{The zero mode}

\section{Discussion}

We have studied particle creation due to a massless bulk spin-$1/2$ field in a 5D warped cosmological braneworld scenario.  We analysed the distinguishing effects of various components of such models, i.e. the warping factor, the on-brane cosmological expansion factor and the dynamic scale of the extra dimension on matter creation. To solve the Dirac equation, we used a simple ansatz which is a simplest possible extension of what one uses for 4D FLRW background. This helps us to compare our results with the four dimensional case. The renormalized energy momentum tensor is derived using the adiabatic regularization method developed in \cite{Ghosh:2015mva,Ghosh:2016epo}. The leading order term of the renormalised energy density, which is of sixth adiabatic order, is explicitly calculated. Below we summarize our findings. 
\begin{itemize}
\item
The leading contribution in energy density for each $q$, where $q$ represents the so-called Kaluza-Klein tower, is proportional to $q^{-2}$. For growing (decaying) warp factor, $q$ takes continuous (discrete) values. Interestingly, $q/b(t)$ acts like the (time-dependent) mass of the projected or on-brane spin-$1/2$ field. Unlike the case for scalar fields, particles in $q=0$ mode are not produced. 

\item
The warping factor determines how much of the created matter density would be localised at different locations along the extra dimension. It is found that a growing warp factor creates a thick brane as such. 
Whereas, in presence of a finite extra dimension, a decaying warp factor pushes matter onto the boundaries creating thin branes.
The presence of warp factor essentially helps matter to accumulate at the minimum of the warp factor. This suggests that as many parallel braneworlds would be formed as the number of minima in the warp factor.

\item The adiabatic series solutions for the Z-S variables play key role in our method of adiabatic regularization. Interestingly, comparison of equations for Z-S variables with 4D case shows that the role of the variables $u_k$ and $\t_k$ is exchanged. An intuitive understanding of these correspondence surely will arise from a clear idea about what $u_k$ and $\t_k$ actually represent, like $s_k$ always represents particle number density in mode $k$. We plan to report more on this in detail elsewhere.

\item Using a toy model for a cosmological brane with a bounce, where $a(t)=\sq{1+(t/t_0)^2}$, we showed that if the extra dimension is expanding it contributes a positive energy density at all times and thus flattens the effect of the bounce on the energy density which in fact becomes negative at late times. On the other hand, a shrinking extra dimension has an opposite effect on the energy density. These and other derivations including the on-brane conformal anomaly suggests that for brane dynamics the ratio $a(t)/b(t)$ plays the role of effective cosmological scale factor.
%%%%%%%%%%%%%%%%%%%%%%%%%%%%%%%
\end{itemize}
%%%%%%%%%%%%%%%%%%%%%%%%%%%%%
In order to study production of on-brane chiral fermions (through dimensional reduction), one may work with a more general ansatz, in chiral representation, for the bulk Dirac field. This may lead to a REMT components with unique features like  non-separability in $t$ and $y$. Particularly, components like $\< T^y_{~y} \> $ and $\< T^0_{~y} \> $ may become non-zero which would suggest matter flow along the extra dimension as such. Further, such analysis may dynamically explain why the universe has a preference for left-handed chirality.
It was shown earlier \cite{Maeda:1983fq,Maeda:1984un}, that the back-reaction from particle production eventually lead to isotropisation which could be crucial as our background metric is essentially anisotropic. It will be interesting to analyse the time scale of such bulk isotropisation process. An initial anisotropy could explain the creation of primordial magnetic fields as well. 
It would also be essential to compare the on-brane matter density with the critical energy density to test the viability of various models. Some of these results may depend on the exact dynamic nature of $a(t)$ and $b(t)$ as well. Further, the background perturbations in the bulk \cite{Gordon:2000pt} may affect particle creation on the brane as well which we did not consider here and should be addressed separately.
We plan to report on these in future.

\section*{Acknowledgments}

This research is supported by a start-up grant awarded by the University Grant Commission, India, with grant number no. F.30-420/2018(BSR). 
Author would like to thank Sayan Kar for helpful discussions and comments. %and the `Visitors Program' of the Centre for Theoretical Studies at the Indian Institute of Technology Kharagpur where the initial part of this work was done. 

\appendix

\section{Useful geometric quantities} \la{ap:geo}

%Following are few useful formulas for cosmological braneworld geometry.

The non-zero affine connections for metric (\ref{eq:metric}):
\be
\Gamma^0_{00} = \Gamma^0_{ii} = \Gamma^i_{i0} = \f{\dot{a}}{a}, ~~ \Gamma^\m_{\m y} = \Gamma^0_{0y} = f'; ~~~~ \m = 0,1,2,3.
\ee
%%%%%%%%%%%%%%%
\be
\Gamma^y_{00} = - \Gamma^y_{ii} = \f{a^2 e^{2f}}{b^2}f' ,~~~~ \Gamma^0_{yy} = \f{b \dot{b}}{a^2 e^{2f}}, ~~~~ \G^y_{0y} = \f{\dot b}{b}.
\ee

%%%%%%%%%%%%%%%
%\be
%\mbox{Ricci scalar:}~~ R = \f{e^{2f}}{a^2} \le(6\f{\ddot{a}}{a} + 4\f{\dot{a}\dot{b}}{ab} + 2 \f{\ddot{b}}{b}\ri) - \f{8f'' + 10 f'^2}{b^2}, \la{eq:R}
%\ee

The non-zero spin connections derived using equation (\ref{eq:spincon}) are given by
\ba
\o_0 &=& \f{e^f\, a\, f'}{2\,b} \g_0\, \g^y  ,\\
\o_i &=& \f{\dot a}{2a} \g_0\, \g^i + \f{e^f\, a\, f'}{2\,b} \g_i\, \g^y , \\
\o_y &=& \f{\dot b}{2b\, e^f} \g_0\, \g^y.
\ea

%%%%%%%%%%%%%%%%%%%%%
%\ba
%R_{ABCD}R^{ABCD} &=& \f{e^{-2f}}{a^4}  \la{eq:Rmn-sq	}
%\ea
%
%
%%%%%%%%%%%%%%%%%%%%%
%\be
%R_{\m\n}R^{\m\n} = 12\le(\f{\dot{a}^4}{a^8} - \f{\dot{a}^2\ddot{a}}{a^7}	+ \f{\ddot{a}^2}{a^6}\ri). \la{eq:Ricci-sq	}
%\ee
%
%%%%%%%%%%%%%%%%%%%%%
%%\be
%%R_{\m\n}R^{\m\n} = 12\le(\f{\dot{a}^4}{a^8} - \f{\dot{a}^2\ddot{a}}{a^7}	+ \f{\ddot{a}^2}{a^6}\ri). \la{eq:Ricci-sq	}
%%\ee
%
%
%\be
%\mbox{Gauss-Bonnet scalar:}~~G = 24\, \f{a\dot{a}^2\ddot{a} - \dot{a}^4}{a^8}.
%\ee
%
%%%%%%%%%%%%%%%%%%
%\be
%\square R = 6 \le(\f{3\ddot{a}^2}{a^6} - \f{6\dot{a}^2\ddot{a}}{a^7} + \f{4\dot{a}\dddot{a}}{a^6} - \f{\ddddot{a}}{a^5}\ri), \la{eq:boxR}
%\ee

\section{WKB solution} \la{ap:wkb}

To find the WKB solution to equations (\ref{eq:h1}) and (\ref{eq:h1}), we restore $\hb$, which implies
\ba
{\ddot h}^I_{kq} - \f{\dot \O}{\O} {\dot h}^I_{kq} + \f{|\O |^2}{\hb^2} h^{I}_{kq} &=& 0, \la{eq:h1ap} \\
{\ddot h}^{II}_{kq} - \f{\dot \O^*}{\O^*} {\dot h}^{II}_{kq} + \f{|\O |^2}{\hb^2} h^{II}_{kq} &=& 0, \la{eq:h2ap}
\ea
Let us assume,
\be
h^{I}_{kq}(t) \sim \exp\le[\int\Big(X(t) + i Y(t)\Big)dt\ri] \la{eq:wkbh1}
\ee
\be
\mbox{where}~~ X(t) = \f{1}{\hbar} \sum_{n=0}^{\infty}\hbar^n X_n(t), ~~Y(t) = \f{1}{\hbar} \sum_{n=0}^{\infty}\hbar^n Y_n(t). \la{eq:XY}
\ee
Putting equation (\ref{eq:wkbh1}) in equation (\ref{eq:h1ap}) and equating the terms of zeroth order in $n$, i.e. terms of the order of $\hb^{-2}$, we get
\ba
X_0^2 - Y_0^2 + 2i\, X_0 Y_0 + |\O|^2 = 0, \la{eq:wkb1}
\ea
which implies
\be
X_0 =0 , ~~~~ Y_0 = \pm |\O|.
\ee
Similarly, solving for the first order in $n$, leads to
\ba 
i \dot{Y}_0 + 2i Y_0X_1 - 2Y_0Y_1 - i \f{\dot \O}{\O} Y_0 &=& 0, \la{eq:wkb2}
\ea
which implies
\be
Y_1 =0 , ~~~~ X_1 = \f{1}{4} \le(\f{\dot \O}{\O} -\f{\dot \O^*}{\O^*}\ri).
\ee
Higher order terms can be neglected in the WKB approximation. 
Thus, equation (\ref{eq:wkbh1}) leads to
\be
h^{I}_{kq} (t) \sim \f{1}{\sq{2}} \le(\f{\O}{~\O^*}\ri)^{\f{1}{4}} \exp\le[i \int |\O| dt\ri]
\ee
Similarly one can solve equation (\ref{eq:h2ap}). %Note that the approximations made above are valid in the adiabatic limit. Therefore equation (\ref{eq:g1g2}) represents the adiabatic vacuum.

\section{Equations for Z-S variables in FLRW Universe}  \la{ap:4D}

The equations for Z-S variables in the four dimensional FLRW Universe are given by \cite{Ghosh:2015mva,Ghosh:2016epo}
\ba
\dot{s}_k &=& F\ u_k, \la{eq:s4D}\\
\dot{u}_k &=& 2F(1-2s_k) - 2 \O_k \t_k, \la{eq:u4D}\\
\dot{\t}_k &=& 2\O_k u_k. \la{eq:t4D}
\ea

\section{$s_q^{(r)}$, $u_q^{(r)}$ and $\t_q^{(r)}$ of different adiabatic orders} \la{ap:sut}

At the first adiabatic order we have,
\be
u_{kq}^{(1)} = \f{F}{|\O|} = \f{qk\dot{c}}{2|\O|^3}. \la{eq:u1} 
\ee
%%%%%%%%%%%%%%%%%%%
Then, using equations (\ref{eq:t^r} - \ref{eq:u^r}) recursively, we get
%Putting equation (\ref{eq:t1}) in equation (\ref{eq:t}) we get the leading term in the adiabatic expansion of $u_k$ which is of order two, 
\be
\t_{kq}^{(2)} = \f{3q^3kc\dot{c}^2}{4|\O|^6} - \f{q k \ddot{c}}{4|\O|^4}.	 \la{eq:t2} 
\ee
%%%%%%%%%%%%%%%%%%%%
%Similarly, by putting equation (\ref{eq:u2}) in equation (\ref{eq:s}) we get the leading term in the adiabatic expansion of $s_k$ which is again of order two, 
\be
s_{kq}^{(2)} = \f{q^2k^2\dot{c}^2}{16|\O|^6}. \la{eq:s2} 
\ee
%%%%%%%%%%%%%%%%%%%%%%%%
%Now putting equation (\ref{eq:s2}) again back in equation (\ref{eq:u}) we get the next to leading term in the adiabatic expansion of $\t_k$ which is of adiabatic order three. This iteration leads to the equations (\ref{eq:u^r}-\ref{eq:t^r}) that give for higher orders:
\be
u_{kq}^{(3)} = \f{5q^3k^3\dot{c}^3}{16|\O|^9} - \f{15q^5kc^2\dot{c}^3}{8|\O|^9} + \f{5q^3k\dot{c}\ddot{c}}{4|\O|^7} - \f{qk\dddot{c}}{8|\O|^5}, \la{eq:u3} 
\ee
%%%%%%%%%%%%%%%%%%
%\begin{widetext}

\ba
 \t_{kq}^{(4)} &=& -\f{35q^3k^5\dot{c}^2\ddot{c}}{32|\O|^{12}} - \f{5q^7kc^5\ddot{c}^2}{8|\O|^{12}} - \f{15q^7kc^5\dot{c}\dddot{c}}{16|\O|^{12}} - \f{5q^3k^5c\ddot{c}^2}{8|\O|^{12}}  %\nn \\ && 
+ \f{105q^5k^3c\dot{c}^4}{32|\O|^{12}}  - \f{15q^3k^5c\dot{c}\dddot{c}}{16|\O|^{12}}  \nn \\
&&  - \f{105q^7kc^3\dot{c}^4}{16|\O|^{12}} - \f{5q^5k^3c^3\ddot{c}^2}{4|\O|^{12}} - \f{15q^5k^3c^3\dot{c}\dddot{c}}{8|\O|^{12}} + \f{qk^7\ddddot{c}}{16|\O|^{12}} %\nn \\  && 
 + \f{q^7kc^6\ddddot{c}}{16|\O|^{12}} \nn \\  &&  
  + \f{175q^5k^3c^2\dot{c}^2\ddot{c}}{32|\O|^{12}} + \f{3q^3k^3c^2\ddddot{c}}{16|\O|^{10}} + \f{105q^7kc^4\dot{c}^2\ddot{c}}{16|\O|^{12}},  \la{eq:t4} 
\ea

%%%%%%%%%%%%%%%%%%%
\ba
s_{kq}^{(4)} &=& \f{21q^4k^4\dot{c}^4}{256|\O|^{12}} - \f{21q^6k^2c^2\dot{c}^4}{64|\O|^{12}} + \f{7q^4k^2c\dot{c}^2\ddot{c}}{32|\O|^{10}} %\nn \\ && 
+ \f{q^2k^2\ddot{c}^2}{64|\O|^8} - \f{q^2k^2\dot{c}\dddot{c}}{32|\O|^8}. \la{eq:s4}
\ea
Higher order terms can be derived in similar way.

%$\left(
%\begin{array}{cc}
% \text{om[1, 1, 5]} & \frac{e^{f(y )} a(t) f'(y )}{4 P(t)} \\
% \text{om[1, 5, 1]} & -\frac{e^{f(y )} a(t) f'(y )}{4 P(t)} \\
% \text{om[2, 1, 2]} & \frac{a'(t)}{4 a(t)} \\
% \text{om[2, 2, 1]} & -\frac{a'(t)}{4 a(t)} \\
% \text{om[2, 2, 5]} & \frac{e^{f(y )} a(t) f'(y )}{4 P(t)} \\
% \text{om[2, 5, 2]} & -\frac{e^{f(y )} a(t) f'(y )}{4 P(t)} \\
% \text{om[3, 1, 3]} & \frac{a'(t)}{4 a(t)} \\
% \text{om[3, 3, 1]} & -\frac{a'(t)}{4 a(t)} \\
% \text{om[3, 3, 5]} & \frac{e^{f(y )} a(t) f'(y )}{4 P(t)} \\
% \text{om[3, 5, 3]} & -\frac{e^{f(y )} a(t) f'(y )}{4 P(t)} \\
% \text{om[4, 1, 4]} & \frac{a'(t)}{4 a(t)} \\
% \text{om[4, 4, 1]} & -\frac{a'(t)}{4 a(t)} \\
% \text{om[4, 4, 5]} & \frac{e^{f(y )} a(t) f'(y )}{4 P(t)} \\
% \text{om[4, 5, 4]} & -\frac{e^{f(y )} a(t) f'(y )}{4 P(t)} \\
% \text{om[5, 1, 5]} & \frac{e^{-f(y )} P'(t)}{4 a(t)} \\
% \text{om[5, 5, 1]} & -\frac{e^{-f(y )} P'(t)}{4 a(t)} \\
%\end{array}
%\right)
%$
%

%\begin{thebibliography}{99}
%
%

\end{document}